\documentclass[achemso,twocolumn,showpacs,superscriptaddress,amsmath]{revtex4}
\usepackage{graphicx}
\usepackage{amssymb}
\usepackage{color}
\usepackage{float}

\begin{document}

\title{Diagrammatic Exciton Basis Theory of the Photophysics of Pentacene Dimers}
\author{Souratosh Khan}
\affiliation{Department of Physics, University of Arizona
Tucson, AZ 85721}
\author{Sumit Mazumdar}
\affiliation{Department of Physics, University of Arizona}
\affiliation{Department of Chemistry and Biochemistry, University of Arizona}
\affiliation{College of Optical Sciences, University of Arizona}
\date{\today}
\begin{abstract}
Covalently linked acene dimers are of interest as candidates for intramolecular singlet fission. 
We report many-electron calculations of the energies and wavefunctions of the optical singlets, 
the lowest triplet exciton and the triplet-triplet biexciton, as well as the final states of 
excited state absorptions from these states in a family of phenyl-linked pentacene dimers. 
While it is difficult to distinguish between the triplet and the triplet-triplet from their transient absorptions 
in the 500-600 nm region, by comparing theoretical transient absorption spectra against 
published and unpublished experimental transient absorptions in the near and mid infrared 
we conclude that the end product of photoexcitation in these particular bipentacenes is 
the bound triplet-triplet and not free triplets. 
We predict additional transient absorptions at even longer wavelengths, beyond 1500 nm, to the equivalent of the classic 2$^1$A$_g^-$ in linear polyenes. 
\end{abstract}
\maketitle
\begin{figure}[H]
\includegraphics[width=3.5in]{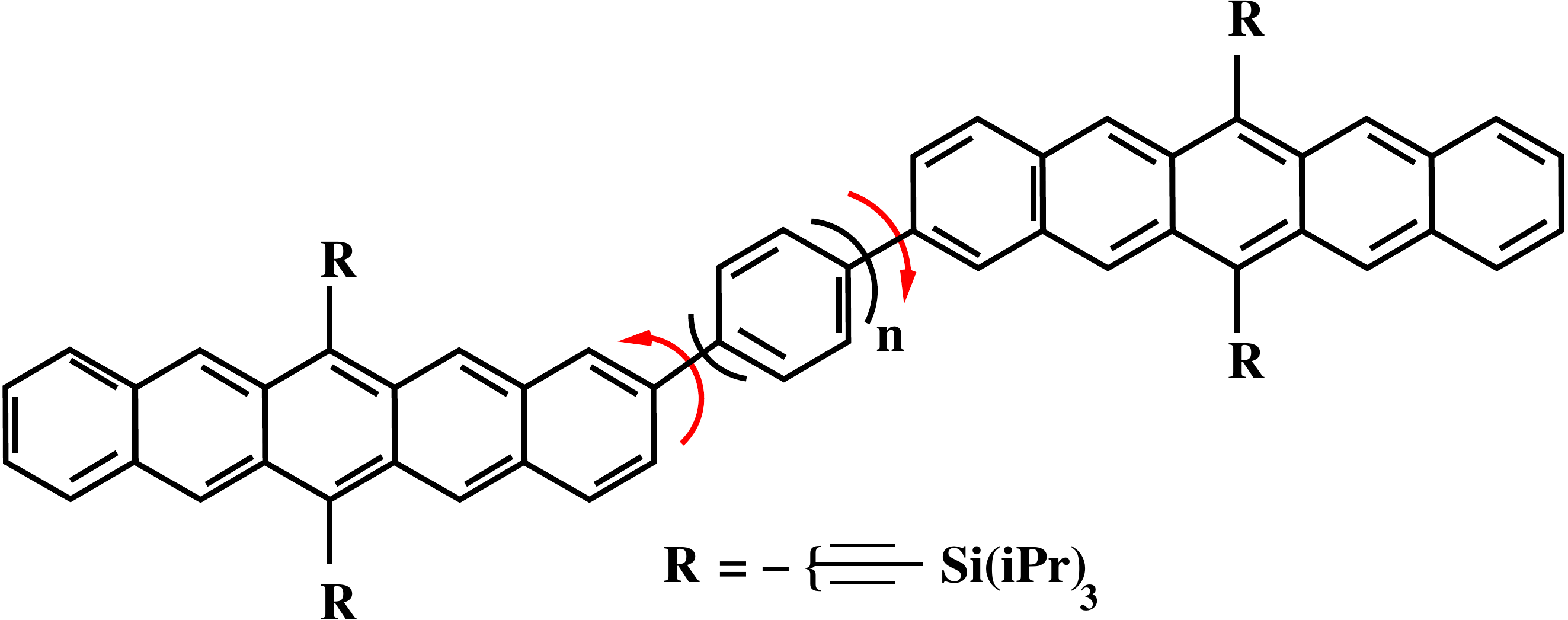}
\caption{BPn dimer, with pentacene monomer molecules linked by n = 0, 1 and 2 phenyl groups. The curved arrows denote rotations about interunit bonds.}
\label{BPn}
\end{figure}

The consequences of strong electron correlations in $\pi$-conjugated systems have been investigated most intensively at the two
extremes of system sizes, (a) small molecules such as linear polyenes, and (b)  
extended systems, such as $\pi$-conjugated polymers and single-walled carbon nanotubes. The occurrence of the
lowest two-photon 2$^1$A$_g^-$ state below the one-photon 1$^1$B$_u^+$ optical state in the former was of strong
interest in the past \cite{Hudson82a,Ramasesha84c,Tavan87a,Schmidt12a}. The  2$^1$A$_g^-$ plays a weak role in the photophysics of most 
extended systems \cite{Guo94a}, where 
the phenomena of interest are exciton formation \cite{Leng94a,Wang05b} and the consequence thereof
on nonlinear optical spectroscopy \cite{Zhao06a}. Understanding the photophysics of discrete but {\it large} molecular systems
of intermediate size poses new challenges \cite{Raghu02a,Aryanpour14a,Aryanpour14b,Chakraborty13a}. We consider members of one such family of large $\pi$-conjugated molecules here, 
dimers of bis(triisopropylsilylethynyl) (TIPS) pentacene, covalently linked by 
0, 1 and 2 phenyl groups (see Fig.~\ref{BPn}) \cite{Sanders15a}. Following the original investigators we will refer to these molecules as 
bipentacenes BPn, with n = 0, 1 and 2, respectively.
The photophysics of these and other bipentacenes \cite{Sanders15a,Zirzlmeier15a,Lukman15a,Fuemmeler16a,Sakuma16a} and related covalently linked dimeric molecular systems \cite{Sanders16a,Korovina16a,Liu15b,Margulies16a}
are of strong current interest 
as candidates for {\it intra}molecular singlet fission (iSF), as we briefly discuss below.

SF is the process \cite{Singh65a} in which an optical spin-singlet exciton S$_1$ dissociates into two spin triplet excitons T$_1$. If 
each photogenerated triplet dissociates with 100\% efficiency at the donor-acceptor interface of an organic solar cell, the photoconductivity
is doubled. Enhanced external quantum efficiencies using SF have been reported for pentacene/C$_{60}$ solar cells \cite{Yoo04a,Congreve13a}. The bulk of the
theoretical \cite{Smith10a,Greyson10a,Kuhlman10a,Zimmerman11a,Smith13a,Beljonne13a,Chan13a,Zeng14a,Yost14a,Aryanpour15a} and experimental \cite{Jundt95a,Thorsmolle09a,Marciniak09a,Chan11a,Wilson13b,Lee13a,Herz15a,Bakulin16a,Pensack16a,Walker13a} literature until now had focused on {\it inter}molecular SF (xSF), in which the two triplets
are generated on neighboring weakly coupled {\it monomer} molecules in a thin film or crystal. 
Recent investigations of dimer molecules \cite{Sanders15a,Zirzlmeier15a,Lukman15a,Fuemmeler16a,Sakuma16a,Sanders16a,Korovina16a,Liu15b,Margulies16a}, in which
the monomers are linked by covalent bonds, is driven by the belief that the stronger coupling between the monomer components will 
give higher SF efficiency.
We present here the results of theoretical investigations of the photophysics of BPn, focusing on electronic
states reached by ground state as well as transient absorption \cite{Sanders15a}. 
We have performed high order
configuration interaction (CI) calculations within the $\pi$-electron Pariser-Parr-Pople (PPP) Hamiltonian \cite{Pariser53a,Pople53a}.
The key difference between earlier theoretical work on bipentacenes \cite{Lukman15a,Zirzlmeier15a,Fuemmeler16a} 
and our work is that we retain much larger
active space, 24 molecular orbitals (MOs) overall, and incorporate CI with up to quadruple excitations from the Hartree-Fock (HF) ground state. 
We thus obtain energies and wavefunctions of all relevant eigenstates from direct CI calculations on PPP Hamiltonians with dimension $\sim$ 10$^6$.
We have investigated the optically accessible spin-singlet states S$_n$ ($n>0$, $n=0$ referring to the ground state), 
the lowest triplet exciton T$_1$, as well as the 
triplet-triplet state $^1$(TT)$_1$ state that is accepted to be the key intermediate in the SF process (here the superscript
refers to the spin multiplicity while the subscript indicates that it is the lowest triplet-triplet state). As in the 2$^1$A$_g^-$ in the polyenes,
the spin triplets in $^1$(TT)$_1$
are quantum-entangled to give an overall spin singlet state, and thus this state
could have been written as S$_n$. Our nomenclature, that labels only one-photon allowed singlet states as S$_n$, 
makes distinguishing between optically allowed and 
dark states simpler (note that this implies that $n$ is not a quantum number). Additionally, barring complete CI calculation which is not possible
even for BP0, different computational methodologies even within
the PPP model will determine different quantum numbers for the same excited eigenstate; our labeling of states
will allow straightforward  comparisons to existing and future theoretical work. 

As in our recent work on pentacene crystals \cite{Khan17a}, we obtain physical pictorial descriptions of S$_1$, T$_1$ and $^1$(TT)$_1$, and also 
report calculations of excited state absorptions (ESAs) from these states in BP0, for direct comparisons of theoretical results 
against experimental ultrafast transient absorption study \cite{Sanders15a}. A primary motivation is to elucidate the key differences 
between xSF and iSF. To this end, we investigate (i) the roles of the phenyl linkers in BP1 and BP2, (ii) the role of intramolecular charge-transfer (CT) 
between the pentacene units, and 
(iii) the difference between ESA from the $^1$(TT)$_1$ and free T$_1$. The last is particularly important, since this difference, if any,
is the only reliable means of determining whether fission is indeed occurring to generate free triplets. 
We are not aware of existing calculations of transient absorptions.
Although our focus is on BPn, our computational methodology can be easily extended to other pentacene dimers, and it is likely that the overall results
are applicable even to other dimeric systems.

{\it \underbar{PPP Hamiltonian \cite{Pariser53a,Pople53a}}.} The PPP Hamiltonian is written as,
\label{PPP_Ham}
\begin{eqnarray}
H=\sum_{\langle ij \rangle,\sigma}t_{ij}(c_{\mu i\sigma}^{\dagger}c_{\mu j\sigma}+c_{j\sigma}^\dagger c_{i\sigma}) \\ 
\nonumber
+ U\sum_{i}n_{i\uparrow}n_{i\downarrow} + \sum_{i<j} V_{ij} (n_i-1)(n_j-1) \\
\nonumber
\end{eqnarray}
\noindent where $c^{\dagger}_{i\sigma}$ creates a $\pi$-electron of spin $\sigma$ on the p$_z$ atomic orbital of carbon (C) atom $i$,
$n_{i\sigma} = c^{\dagger}_{i\sigma}c_{i\sigma}$ is the number of electrons of spin $\sigma$ on atom $i$,
and $n_{i}=\sum_{\sigma} n_{i\sigma}$ is the total number of electrons on the atom.
We consider hopping integrals $t_{ij}$ only between nearest neighbor (n.n.) C atoms $i$ and $j$ ($\langle .. \rangle$ denote n.n.),
$U$ is the onsite Coulomb repulsion between two electrons occupying the same p$_z$ orbital, and $V_{ij}$ is the long range interatomic
Coulomb interaction, respectively. 

As mentioned above, the $^1$(TT)$_1$ state plays a very important role in SF. The $^1$(TT)$_1$ state 
is dominated by many-electron configurations that are doubly excited from the HF ground state, and competes with
single excitations that are CT within  our basis space (see below). A balanced description requires treating all one electron - one hole 
(1e-1h) and two electron - two hole (2e-2h) excitations 
on equal footing, which actually requires incorporating CI with up to
quadruple (4e-4h) excitations \cite{Tavan87a,Schmidt12a}. This is because 2e-2h excitations are coupled to both the ground state configuration as well as 4e-4h
excitations by the many-electron components of the PPP
Hamiltonian \cite{Tavan79a}. Since the number of ne-nh configurations increases steeply with n for large molecules, including CI with  
quadruple excitations is difficult within first principles approaches for systems containing more than 8-10 electrons, and would be impossible for BPn without
severely restricting the active space of MOs about the chemical potential. Overly restricted active space, however, 
gives unbalanced descriptions of CT versus $^1$(TT)$_1$, and also makes calculations of ESAs, necessary to compare
against transient absorption measurements, impossible. Thus the correct determination of the energy and wavefunction of $^1$(TT)$_1$ 
requires high order CI calculations over a large active space of MOs. Below we report calculations using the multiple reference singles and doubles CI (MRSDCI),
which incoporates CI with the most dominant 1e-1h to 4e-4h configurations \cite{Tavan87a,Aryanpour15a,Khan17a}, 
over active spaces of 12 bonding and 12 antibonding MOs. This active space allows including the benzene-derived MOs in exciton basis
calculations (see below). 
\begin{figure}[H]
\includegraphics[width=3.5in]{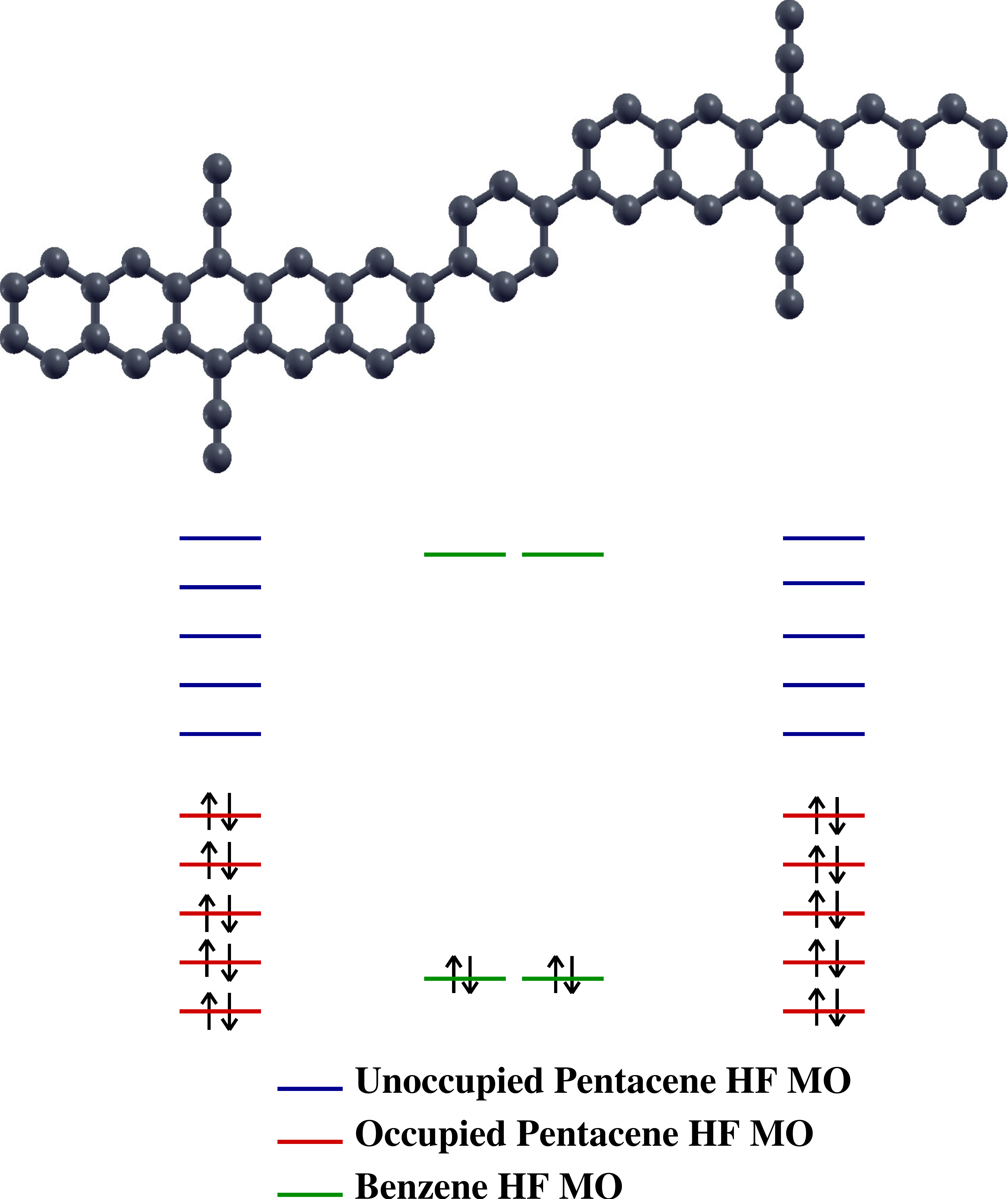}
\caption{Hartree-Fock MOs localized on the individual pentacene chromophores and the phenyl linker in BP1 that are included in the PPP-MRSDCI calculation
for BP1. The locations of the MOs show their relative energies schematically.}
\label{MOs}
\end{figure}

{\it \underbar{Exciton Basis}.} The PPP Hamiltonian, as written in Eq.~1, is not suitable for our purpose, as our goal is to
distinguish between excitations localized within individual pentacene segments and those that are delocalized over the
entire bipentacene molecule. This requires the use of MOs that are localized on individual components \cite{Chandross99a,Aryanpour15a,Khan17a}. 
We rewrite Eq.~1 as
\begin{equation}
H = H_{intra} + H_{inter} \\
\label{PPP_split}
\end{equation}
where $H_{intra} = \sum_i H_{intra}^i$ is the sum of PPP Hamiltonians describing individual molecular units $i$ (the two pentacene segments and the phenyl linkers)
and  $H_{inter} = \frac{1}{2} \sum_{i,j} H_{inter}^{i,j}$ includes the remaining terms in Eq.~1, consisting of hopping integrals $t_{ij}$ and 
Coulomb interactions $V_{ij}$ between C atoms belonging to the
different units \cite{Chandross99a}. The calculations are now done in multiple stages. First, $H_{intra}^i$ are solved at the HF level to give 
MOs localized on individual units $i$ (see Fig.~\ref{MOs}). The basis space many-electron configurations are then constructed by filling in the localized MOs with
appropriate numbers of {\it delocalized} electrons, retaining ground state (all bonding MOs filled and antibonding MOs empty) as well as excited (partially occupied
bonding and antibonding MOs) configurations. The configurations
can be {\it neutral} 
(number of $\pi$-electrons same as number of C atoms within each unit) and {\it ionic} (units positively
and negatively charged). 
A schematic of the MOs
retained is shown in Fig.~\ref{MOs} for the specific case of BP1. Note that our active space includes the degenerate pairs of highest bonding and lowest antibonding MOs 
belonging to the benzene groups.
The exciton basis is diagrammatic, allowing pictorial descriptions of all eigenstates as superpositions of the most dominant many-electron configurations \cite{Chandross99a,Khan17a}. 
In what follows, we will emphasize energies as well as wavefunctions.

{\it \underbar{Multiple Reference CI}.} The MRSDCI procedure incorporates the most dominant ne-nh
configurations (n=1-4) that describe each targeted excited state \cite{Tavan87a,Aryanpour15a,Khan17a}. 
The calculation for each target eigenstate is done iteratively,
with each iteration consisting of two stages. In the first stage we perform a double-CI calculation on a basis space of N$_{ref}$ 
1e-1h and 2e-2h configurations that best describe the targeted eigenstate. In the second stage we 
apply the Hamiltonian ($H_{intra} + H_{inter}$) on the  N$_{ref}$ reference configurations. This generates 3e-3h and 4e-4h configurations, of which we 
retain the most dominant ones to give the larger Hamiltonian matrix of dimension N$_{total}$. The larger Hamiltonian matrix usually also contains new 1e-1h and 2e-2h
excited  
configurations that were not among the original N$_{ref}$ reference configurations, but that are coupled to the 3e-3h and 4e-4h configurations
reached from them. N$_{ref}$ is now updated by incorporating the new 1e-1h and 2e-2h configurations.
The entire procedure is repeated with updated N$_{ref}$ configurations to reach a new larger Hamiltonian with updated N$_{total}$,
until the convergence criterion is reached. N$_{ref}$ and N$_{total}$ can exceed 10$^2$ and 10$^6$, respectively (see Supporting Information, sections I, III, IV).
Although the targeted wavefunctions therefore are superpositions of very large number of configurations, in most cases they can be described  pictorially using the
most dominant many-electron configurations. 

{\it \underbar{Parametrization of PPP Hamiltonian}.} For the hopping integrals $t_{ij}$ we choose $-$2.4 eV and $-$2.2 eV for the peripheral and internal phenyl
bonds (see Fig.~\ref{BPn}) \cite{Aryanpour15a}, and $-$3.0 eV for the carbon-carbon triple bond \cite{Ducasse82a} of the TIPS group, respectively. 
Steric hindrance leads to rotations of the molecular units about the single bonds linking them 
\cite{Annibale73a}.
The interunit hopping integrals are taken to be $-$2.2 cos($\theta$) eV \cite{Ramasesha91a}, where $\theta$ is the angle between any two consecutive molecular planes,
which is a variable in our calculations. 
\begin{table}[H]
\centering
\setlength{\tabcolsep}{8pt}
\caption{Experimental versus calculated energies and energy differences in eV for the TIPS-pentacene monomer for two different parameter sets. T$_4$ is the
monomer triplet state to which excited state absorption is allowed. a, b and c correspond to references 53, 54 and 55 respectively.}
\begin{tabular}{l c c c}
\hline\hline
          & Expt     &      U = 6.7 eV       &       U = 7.7 eV \\[0.5ex]
                    &          &      $\kappa$ = 1.0   &       $\kappa$ = 1.3 \\[0.5ex]
\hline

$S_{1}$ & 1.81$^a$, 1.9$^b$ & 1.88 & 2.22   \\ [0.5ex]

$T_{1}$ & 0.86$^c$ & 0.90  & 0.89  \\ [0.5ex]

$T_{4}-T_{1}$ & 2.46$^b$ & 2.1  & 2.39  \\ [0.5ex]

\hline
\end{tabular}
\label{monomer}
\end{table}
\begin{figure*}
\includegraphics[width=\textwidth,height=13cm]{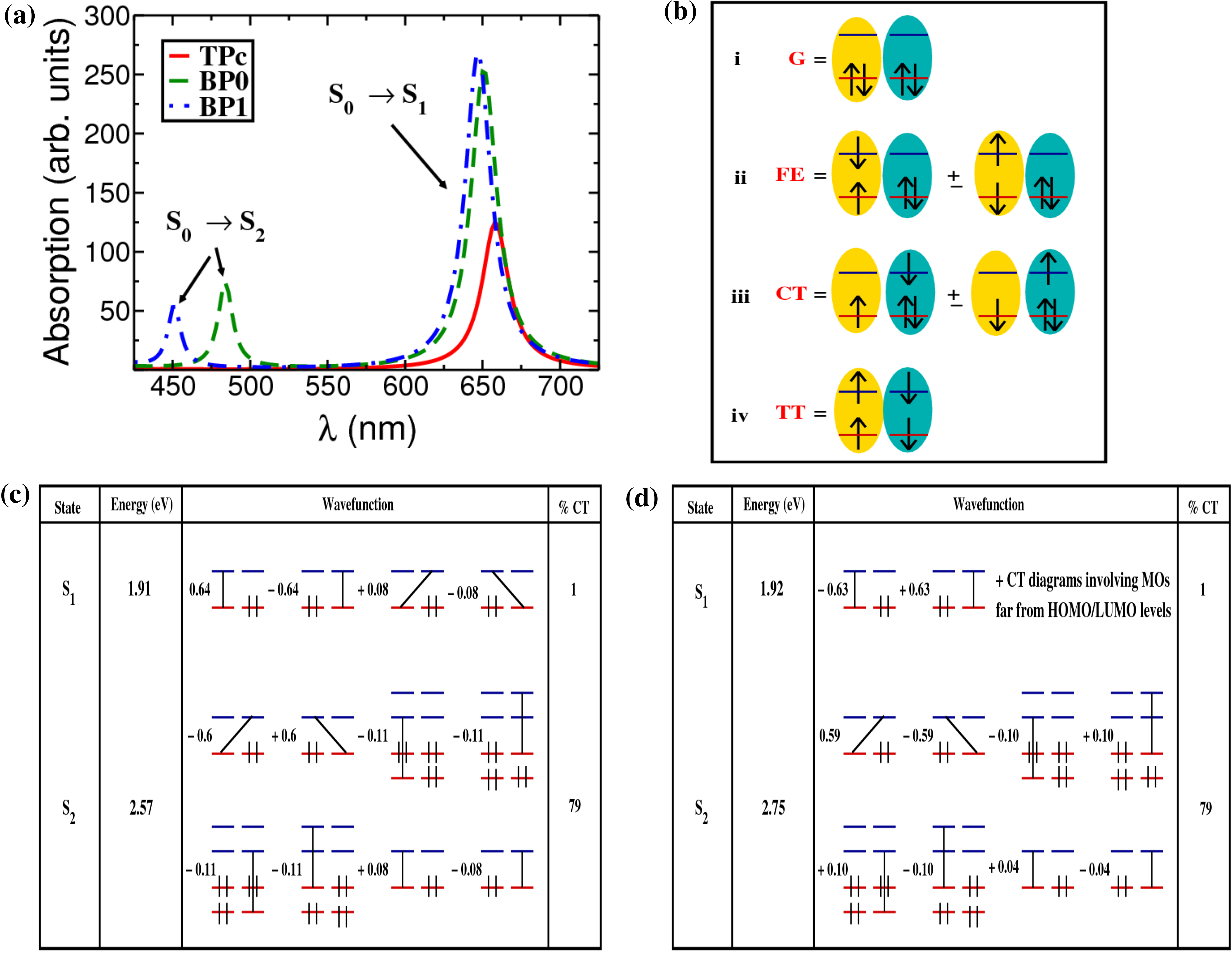}
\caption{(a) Calculated ground state absorption spectra of TIPS pentacene monomer (solid red curve), BP0 (green dashed curve),
and BP1 (blue dashed-dotted curve) for $\theta$ = 0$^{\circ}$
and $U$ = 6.7 eV, $\kappa$ = 1.0. The weak blue shift (instead of red shift \cite{Sanders15a}) of BP0 and BP1 absorption band in the low energy region, relative to the 
monomer absorption indicates that even our large active space is not sufficient for a quantitaively accurate calculation (see section III, Supporting Information).
The final states of the absorptions are labeled as
S$_1$ and S$_2$, respectively, in increasing order of energy. (b) Pictorial representations and nomenclatures of the
simplest many-electron configurations with excitations across the HOMO-LUMO gap. Here the horizontal lines are the HOMO and LUMO
of individual pentacene components, whose electron occupancies are indicated in the figures.
G = ``ground state'' configuration, FE = Frenkel exciton, CT denotes schematically configurations with intramolecular charge-transfer
between the pentacene components and TT denotes a triplet-triplet 2e-2h excitation.
(c) Diagrammatic exciton basis contributions (see text) to the normalized S$_1$ and S$_2$ wavefunctions of BP0. Lines
connecting the bonding and antibonding MOs represent spin-singlet excitations. S$_1$ is predominantly Frenkel exciton, while S$_2$
consists predominantly of
CT excitations, with smaller contributions from higher energy Frenkel excitations.
(d) Same as (c) for BP1, where the MOs belong to the pentacene monomers only (see text).}
\label{grstabs}
\end{figure*}
We parametrize the Coulomb interactions as $V_{ij}=U/\kappa\sqrt{1+0.6117 R_{ij}^2}$, where $R_{ij}$ is the distance in $\mathring{\textrm{A}}$
between C atoms $i$ and $j$ and $\kappa$ is an effective dielectric constant \cite{Chandross97a}.
The Coulomb interaction parameters are determined by $U$ and $\kappa$ only.
As in reference \onlinecite{Khan17a}  we choose our parameters based on fitting the TIPS pentacene monomer singlet and triplet excitation energies. 
{\it We then perform the calculations for the dimer using the same parameters.} We find that while the
energies of S$_1$ (the lowest one-photon optical state) and T$_1$ in the monomer can be fit with a single set of $U$ and $\kappa$ ($U=6.7$ eV, $\kappa=1.0$), 
it is difficult to fit the triplet ESA energy with the same set of
parameters. We have therefore performed our calculations for a narrow range of parameters $U=6.0 - 8.0$ eV, $\kappa=1.0-1.5$. By trial and error we find that
the triplet ESA energy is fit best with slightly larger $U=7.7$ eV and
$\kappa=1.3$ (see also reference \onlinecite{Khan17a}). Table I shows the experimental and computed monomer energies of S$_{1}$, T$_{1}$ and 
the triplet excited state absorption energy (the final state of triplet ESA is labeled T$_4$ instead of T$_2$ in view of what follows). In the following, we therefore show results for
ground state absorption and ESA spectra in the singlet subspace for $U=6.7$ eV, $\kappa=1.0$, and triplet and triplet-triplet ESAs for $U=7.7$ eV, $\kappa=1.3$.
This is for quantitative comparisons to experimental transient absorption spectra only. Our principal goal is to understand the initial and final {\it wavefunctions}
of the ESAs, which we find to be dominated by the same many-electron configurations for these two sets of parameters (see Supporting Information; the differences
in the wavefunctions come largely from the nondominant components). All wavefunctions 
below are shown for the same parameter set $U=6.7$ eV, $\kappa=1.0$, even as the ESA spectra are for different parameters.

{\it \underbar{Energies}.} In Table~II we have given our calculated energies of S$_1$, T$_1$ and $^1$(TT)$_1$ for
n = 0 and 1, for $U=6.7$ eV, $\kappa=1$. Note that in both cases $^1$(TT)$_1$ and S$_1$ are quasidegenerate from direct CI calculations.
The quasidegeneracy of these states arises as a consequence of electron-electron Coulomb correlations alone, as in the case of the  2$^1$A$_g^-$ in the 
polyenes \cite{Hudson82a,Ramasesha84c,Tavan87a,Schmidt12a} and does not require incorporation of electron-molecular vibration interactions.
\begin{table}[h]
\centering
\setlength{\tabcolsep}{12pt}
\caption{Calculated energies (in eV) of the two lowest one-photon optical states, the lowest triplet and triplet-triplet states of BPn for n = 0, 1 and $U$ = 6.7 eV, $\kappa$ = 1.0.}
\begin{tabular}{l c c c c}
\hline\hline
  Compound    &  S$_1$   &   S$_2$     &     T$_1$   &    $^1$(TT)$_1$ \\[0.5ex]
\hline

BP0 & 1.91 & 2.57 & 0.98 & 1.9 \\ [0.5ex]

BP1 & 1.92 & 2.75 & 1.03 & 1.96 \\ [0.5ex]


\hline
\end{tabular}
\label{dimer}
\end{table}

{\it \underbar{Ground state absorption.}} In Fig.~\ref{grstabs}(a) we show our calculated ground state absorption spectra for TIPS pentacene monomer,
BP0 and BP1 for
$U$ = 6.7 eV, $\kappa$ = 1.0 and $\theta=0^{\circ}$. The calculations cannot be done for BP2 while retaining the same active space. We have therefore
performed calculations for a modified BP2 where the monomer is pentacene instead of TIPS pentacene. Because this shifts the monomer absorption, we have shown
the monomer and modified BP2 absorption spectra in Appendix A (see Fig.~A.1). The two absorption spectra of Fig.~\ref{grstabs}(a) and Fig.~A.1 
are qualitatively similar.  
The lower energy absorption to S$_1$  in Fig.~\ref{grstabs}(a) is at the same energy as the monomer absorption, in agreement
with experiments \cite{Sanders15a}, and is polarized along the short axis of the pentacenes. Experiments also find a set of higher energy absorptions for BPn that were not reproduced in  
the density functional theory based calculations in reference \onlinecite{Sanders15a}. The PPP calculations, with $U$ and $\kappa$ parameters obtained from 
fitting monomer absorption, find these absorptions at almost the same wavelengths (450-500 nm as in the experiment, see Fig.~\ref{grstabs}(a)) where the
experimental high energy absorptions are found (see Fig.~2 in reference \onlinecite{Sanders15a}). We have labeled the final state of the higher energy ground state
absorption S$_2$. This absorption is polarized along the long axis of the dimer molecule. 

The weak blue shift (instead of the experimentally observed 
red shift \cite{Sanders15a}) 
of the calculated absorption to S$_1$ in BP0 and BP1, relative to
the TIPS monomer, and the relatively weak intensities of the transition to S$_2$ are both consequences of our choosing an active space that is less than complete. One-photon
absorptions are predominantly 1e-1h in character, and retaining the complete 1e-1h space in a singles-CI calculations do give the red shift of S$_1$ absorption
and a larger intensity of the S$_2$ absorption (see section III of Supporting Information). 

In Fig.~\ref{grstabs}(b) we define 
the fundamental configurations that dominate most of the wavefunctions we
will discuss. In what follows, we will use basis functions that use the total spin $S$ representation, with a bond between MOs representing a spin singlet superposition
of a pair of $S_z=0$ configurations, where $S_z$ is the $z$-component of $S$, and an arrow representing a spin triplet bond which is a superposition of all three
triplet configurations. In many current theoretical works, only these funadamental configurations and a few others are retained in the CI calculations.
We re-emphasize that such extremely small bases are unsuitable for obtaining an accurate description of the true triplet-triplet or obtaining its ESAs. 

In Fig.~\ref{grstabs}(c) we show the most dominant diagrammatic exciton basis contributions to many-electron eigenstates S$_1$ and S$_2$ of BP0. 
The diagrams consist of spin singlet
excitations across the energy gap between the highest occupied and lowest unoccupied MOs (HOMO and LUMO). Where necessary, we have shown the next
lower and higher bonding and antibonding MOs (HOMO-1 and LUMO+1), respectively. We note that even as the dominant configurations are 1e-1h excitations across the HOMO-LUMO gap, 
the sum of the squares of the normalized
coefficients is very far from 1. This is because the {\it number} of higher excitations is huge, and even as individually they make small contributions, their
overall contribution is nonnegligible. 
The same higher order excitations, {\it taken together}, lower the energy of $^1$(TT)$_1$ relative to S$_1$.
The last column in the figure gives the degree of CT between the pentacene components, which is obtained from the summing over the relative weights
of the CT configurations in the wavefunctions multiplied by their ionicities. We note that S$_1$ in BP0 is almost overwhelmingly Frenkel exciton in character: 
the state is a quantum mechanical superposition
of bound excitons on the two units, with little CT contribution. This description of S$_1$ as excitons delocalized over the two pentacene units
agrees with earlier theoretical result \cite{Fuemmeler16a}. 
Simultaneously, S$_2$ is overwhelmingly CT, with much smaller contributions from higher energy intramolecular excitations (HOMO-1 $\rightarrow$ HOMO, LUMO $\rightarrow$ LUMO+1). 
The nearly complete separation of Frenkel versus CT contributions is drastically
different from what is observed in pentacene crystal, where the lowest optical state is nearly 50\% CT \cite{Aryanpour15a,Khan17a}. 
\begin{figure}[H]
\includegraphics[width=2.5 in]{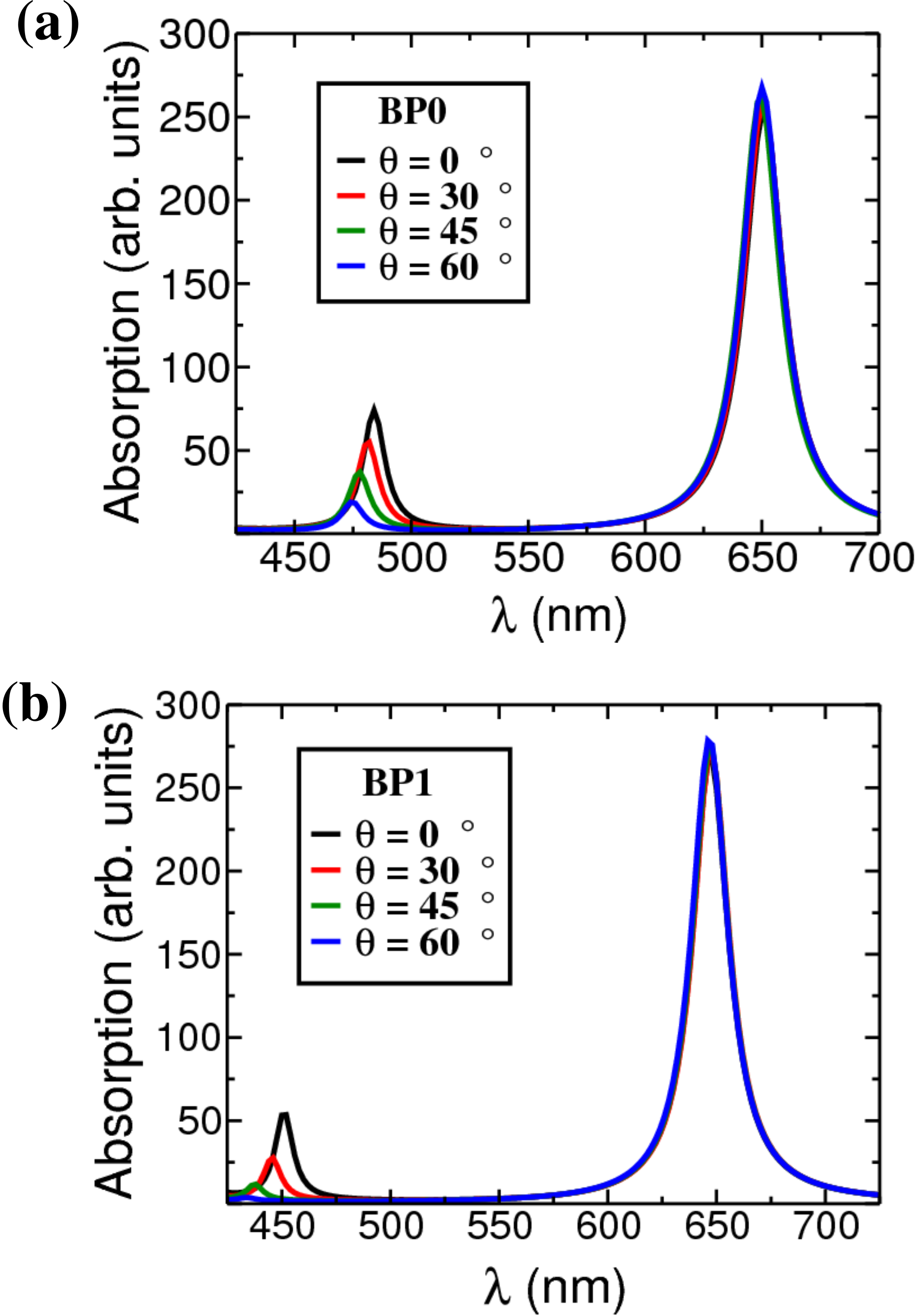}
\caption{Calculated ground state absorption spectra over a range of dihedral angles for (a) BP0 and (b) BP1 with $U$ = 6.7 eV and $\kappa$ = 1.0.}
\label{theta}
\end{figure}
\begin{figure*}
\includegraphics[width=\textwidth,height=7cm]{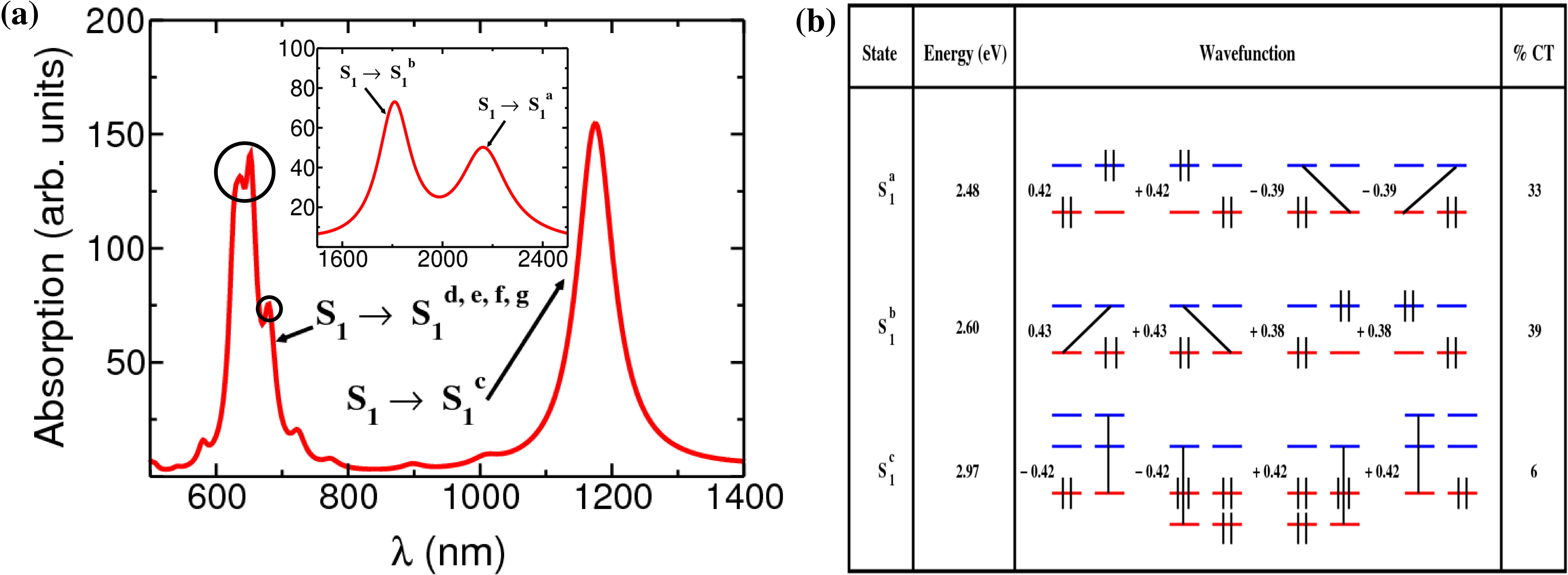}
\caption{(a) Calculated singlet ESA spectrum for $U$ = 6.7 eV and $\kappa=1.0$. The absorption in the visible range is composed of multiple
transitions. The inset shows the calculated ESA in the mid infrared (b) Diagrammatic exciton basis contributions to the final states of singlet PA in the
infrared in BP0 for $U$ = 6.7 eV and $\kappa=1.0$. S$_1^a$ and S$_1^b$ are the equivalents of 2$^{1}$A$_g^-$ in polyenes. The wavefunctions corresponding to the final states of the visible absorption are given in
section IV of the Supporting Information.}
\label{singletPA}
\end{figure*}
\begin{figure*}[t]
\includegraphics[width=\textwidth,height=9cm]{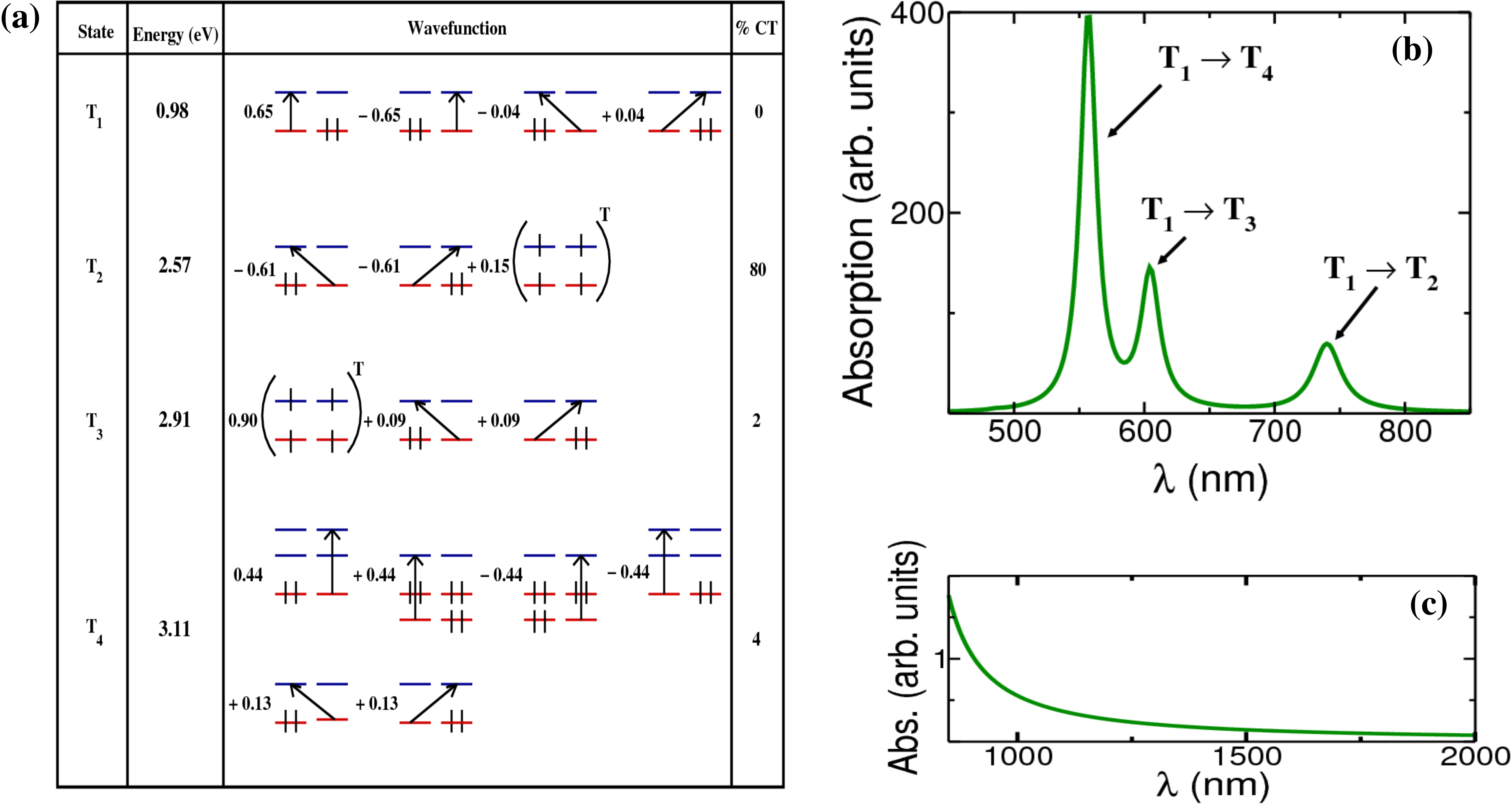}
\caption{(a) Diagrammatic exciton basis contributions to T$_1$ and the final states of triplet PA in BP0 for $U$ = 6.7 eV and $\kappa=1.0$. Here the
arrow represents a spin triplet excitation (see Section IV of Supporting Information). The state
T$_3$ is peculiar to the dimer and is absent in the monomer. The 2e-2h excitation with the superscript `T' is a spin triplet, as opposed to triplet-triplet
which is also 2e-2h. (b) Calculated triplet ESA spectrum in BP0 for $U$ = 7.7 eV and $\kappa$ = 1.3. (c) Absence of
free triplet absorption in the mid infrared.}
\label{tripletPA}
\end{figure*}
The  dominant contributions to the S$_1$ and S$_2$ wavefunctions in BP1 are shown in Fig.~\ref{grstabs}(d).
Our calculations find negligible contribution of the benzene MOs to the excited states: the bonding (antibonding) benzene MOs remain 
almost fully occupied (unocupied)
in both S$_1$ and S$_2$. 
In Fig.~\ref{grstabs}(d) we have therefore included only the MOs of the pentacene units. The energy of S$_2$ is slightly
higher in BP1 (in agreement with experiment \cite{Sanders15a}), but the S$_1$ and S$_2$ wavefunctions in BP0 and BP1 are practically identical. 
In Appendix A we have shown the S$_1$ and S$_2$ wavefunctions for modified BP2. Once again, the benzene-derived  MOs play insignificant 
role, in spite of two phenyl linkers now, and the wavefunctions are nearly pure Frenkel and CT in character.

In Figs.~\ref{theta}(a) and (b) we have shown the calculated absorption spectra as a function of $\theta$ for BP0 and BP1, respectively.
Not surprisingly, molecular rotation leaves the absorption energy as well as intensity in the long wavelength region of monomer absorption unaffected. 
In the shorter wavelength region the primary
consequence is the reduced intensity of the CT absorption, but the absorption energy is affected weakly. Our results are in qualitative agreement with the experimental observation 
of reduced intensities of the shorter wavelength absorption in substituted BP0 compounds in which the dihedral angle between the BP0 units are larger \cite{Fuemmeler16a}.  
In what follows we will mostly show computational results for $\theta=0^{\circ}$, as we will be primarily interested in the lower energy region.
\begin{figure*}[t]
\includegraphics[width=\textwidth,height=6cm]{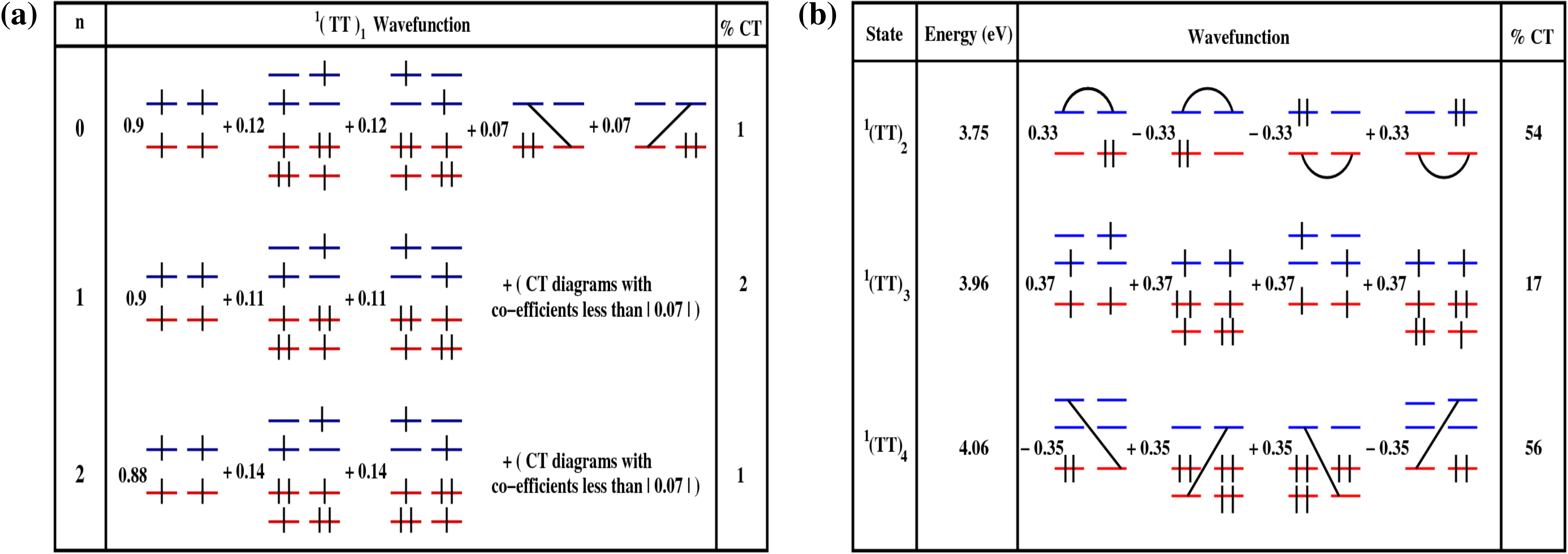}
\caption{(a) Calculated $^{1}$(TT)$_{1}$ wavefunctions for BP0, BP1, BP2 with $U$ = 6.7 eV and $\kappa$ = 1.0 and $\theta$ = 0 $^{\circ}$. (b) The
final states of the ESA from the $^{1}$(TT)$_{1}$ in the visible and near infrared region. There is an additional ESA to S$_2$ in mid infrared.}
\label{triplettriplet}
\end{figure*}

{\it \underbar{Singlet excited state absorption}.} We have calculated singlet ESA from S$_1$, using also the MRSDCI procedure. Fig.~\ref{singletPA}(a)
shows the ESA spectrum over a wavelength region that is much broader than the wavelength region over which experimental singlet photoinduced absorption (PA)
spectrum is shown in reference \onlinecite{Sanders15a}. The motivations behind the calculations of the ESA spectrum over the extended wavelength region 
are to understand a transient absorption that has been
measured subsequently \cite{Sfeir}, as well as to make prediction regarding a wavelength region not yet reached experimentally (both are
discussed in section IV of Supporting Information). In Fig.~\ref{singletPA}(b) we have given the final state wavefunctions
S$_1^a$, S$_1^b$ and S$_1^c$, corresponding to the transitions at wavelengths beyond 1000 nm. 
As seen in Fig.~\ref{singletPA}(b), the transition to S$_1^c$ is ``monomeric'' in character, with the final state wavefunction a superposition of       
configurations in which there has occurred HOMO$-$1 $\to$ HOMO or LUMO $\to$ LUMO+1 transitions. The final states S$_1^a$ and S$_1^b$ at much lower energy
are very different in character; transitions to these states involve ``intermonomer'' CT between the pentacene units as well as double excitations (HOMO, HOMO $\rightarrow$ LUMO, LUMO) within the monomer. {\it These eigenstates are the exact equivalents of the 2$^1$A$_g^-$ in polyenes \cite{Chandross99a}.}  

We find that the shorter wavelength transitions in the visible 500-600 nm region
are to multiple final states and no simple characterization of them is possible. We therefore have shown these final state wavefunctions in Supporting Information
section IV.

Our existing computational facilities do not permit accurate ESA calculations for BP1 and BP2 in the high energy regime. This would require the inclusion of 
even larger active space which would make the computation very expensive. 
Triplet-triplet ESA in the low energy region however can be calculated with 24 active MOS (see section IV of Supporting Information).

{\it \underbar{Triplet exciton and triplet ESA}.} In Figs.~\ref{tripletPA}(a) we show the calculated wavefunctions for the lowest triplet exciton T$_1$ in
BP0 along with the final states of triplet ESA. The subscripts on the higher energy triplet states do not refer to true quantum
numbers but just the order of allowed ESAs. The Coulomb parameters are $U=6.7$ eV, $\kappa=1.0$ for consistent comparisons to the wavefunctions in the singlet subspace. 
As with the singlet excitations, the benzene MOs make very little contribution to the wavefunctions at this energy range. 
Exactly as S$_1$, we find that the T$_1$ Frenkel exciton is delocalized over both units. 

The wavelength-dependent triplet ESA spectrum of BP0 is shown in Fig.~\ref{tripletPA}(b)-(c) for $U=7.7$ eV, $\kappa=1.3$. Section IV of the Supporting Information shows the
triplet wavefunctions of BP0 for this set of parameters. There is little difference between the wavefunctions for the two sets of parameters. 
From Figs.~\ref{tripletPA}(a) and (b) we see that 
the triplet ESA in the wavelength region 550-600 nm actually consists of two distinct absorptions T$_1 \to$ T$_3$ and T$_1 \to$ T$_4$. 
The stronger T$_1 \to$ T$_4$ is ``monomeric''; it is predominantly a superposition of intra-pentacene monomer
LUMO $\to$ LUMO+1 electron excitation and HOMO $\to$ HOMO$-$1 hole 
excitation. Note that the absolute energies of T$_4$ and  S$_1^c$ are close, as should be expected from their wavefunctions. The final state of the weaker T$_1 \to$ T$_3$ absorption involves {\it both} monomers, with a second HOMO $\to$ LUMO excitation in the 
previously unexcited
monomer leading to a 2e-2h excitation. 

In addition to the ESA in the monomer region we find 
a weaker absorption T$_1 \to$ T$_2$ in the infrared region, where the T$_2$ wavefunction consists of inter-pentacene CT components. 
But for the T$_1 \to$ T$_3$ absorption in BP0, the triplet ESA
spectrum is reminescent of that in the pentacene monomer crystal, where for favorable orientations of the monomers there occurs a monomer absorption at short
wavelength and a CT absorption in the near infrared \cite{Khan17a}, with the strength of the latter depending on t$_{inter}$. Experimentally, it is likely that the T$_1 \to$ T$_3$ and T$_1 \to$ T$_4$
absorptions will be overlapping. Finally, as shown in Fig.~\ref{tripletPA}(c), there is no triplet ESA in the longer wavelength ($> 800$ nm) region, a point we will return to.
We have not shown the T1 wavefunctions for BP1 and BP2, which are nearly identical to that of BP0.
As with the singlet excitations, the benzene MOs do not contribute to the wavefunctions at this energy range.

{\it \underbar{The triplet-triplet states}.} We performed MRSDCI calculations for $^1$(TT)$_1$ for both BP0 and BP1, and the modified BP2. In all cases, we found
$^1$(TT)$_1$ and the corresponding S$_1$ to be quasidegenerate (see Table II and also section IV of Supporting Information).
As with S$_1$, S$_2$ and T$_1$, we find that the benzene MOs play a  weak role in the $^1$(TT)$_1$ wavefunctions of BP1 and BP2, in spite of the ``bimolecular''
2e-2h character of $^1$(TT)$_1$.
Fig.~\ref{triplettriplet}(a) shows the diagrammatic exciton basis wavefunctions of the $^1$(TT)$_1$ states for BP0, BP1 and the modified BP2. In all cases in addition
to the lowest triplet-triplet configuration there occur nonnegligible contributions from higher triplet-triplet 2e-2h configurations.
CT contribution to the wavefunctions is negligible in all cases, 
unlike in the pentacene crystal, where CT configurations make significant contribution \cite{Khan17a} to $^1$(TT)$_1$.
\begin{figure}[h]
\includegraphics[width=3.0 in]{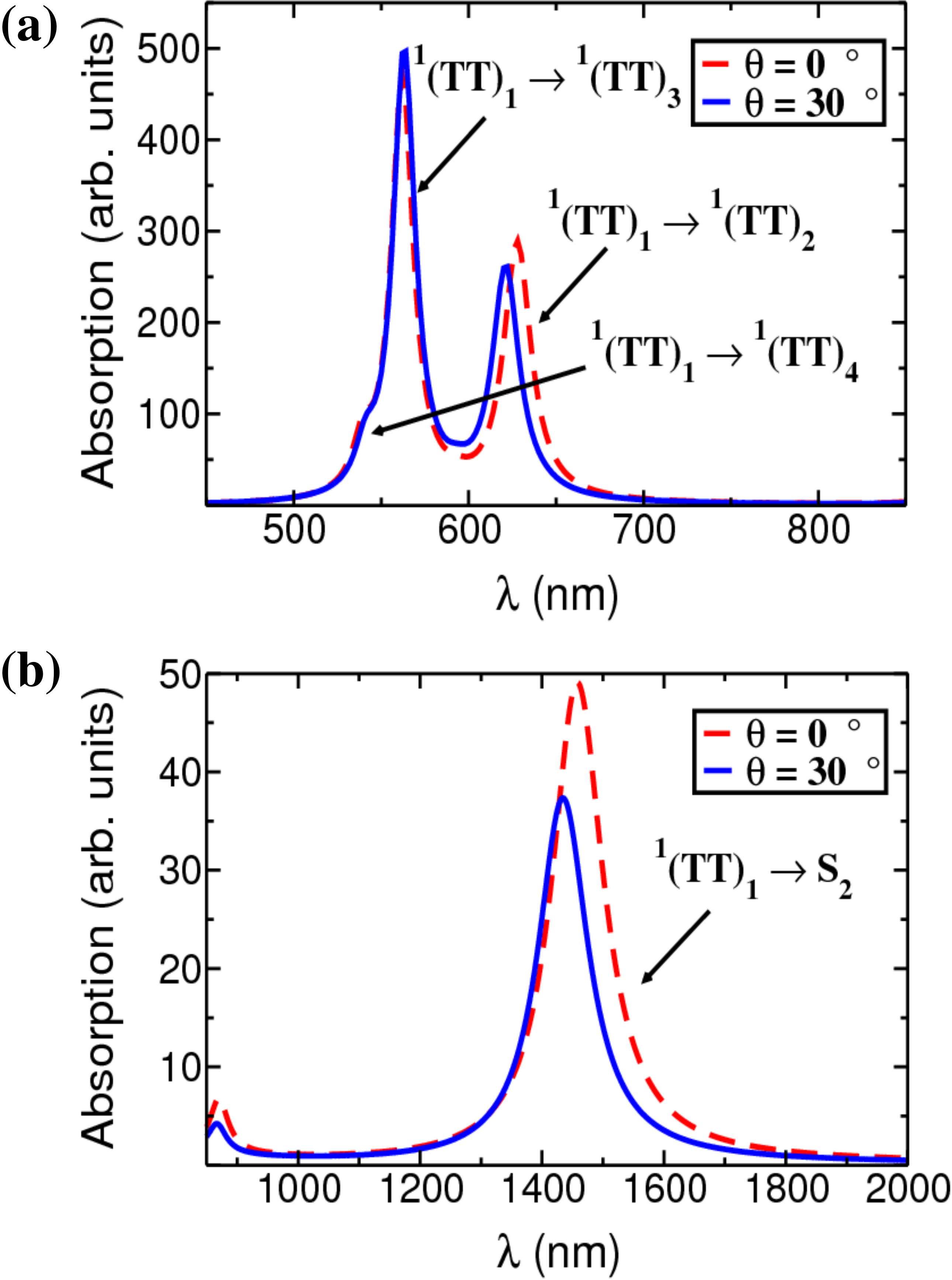}
\caption{Calculated triplet-triplet ESA in BP0 for $\theta$ = 0$^\circ$ $\&$ 30$^\circ$ and $U$ = 7.7 eV, $\kappa$ = 1.3 in (a) visible and (b) mid infrared.}
\label{TTPA}
\end{figure}

{\it \underbar{Triple-triplet excited state absorption}.} Fig.~\ref{triplettriplet}(b) shows the exciton basis wavefunctions of the final
states of the ESAs from $^1$(TT)$_1$ in BP0, while Figs.~\ref{TTPA}(a)  and (b) show the 
corresponding calculated ESA spectrum. Note that the $^1$(TT)$_1 \to$ $^1$(TT)$_3$ transition is nearly identical in character to the monomeric T$_1$ $\to$ T$_4$ transition,
in that both consist of LUMO $\to$ LUMO+1 and HOMO$-$1 $\to$ HOMO transitions. Hence this triplet-triplet ESA and the free triplet transition both occur in the same energy region.
Note also that the final states in Fig.~\ref{triplettriplet}(b) are not all triplet-triplet in character; they are labeled as such for
convenience only, as discussed before. $^1$(TT)$_2$ is a 2e-2h CT state, that is reached from the fundamental triplet-triplet configuration of $^1$(TT)$_1$ by intermonomer
CT. This CT absorption 
occurs at shorter wavelength (higher energy) than the T$_1$ $\to$ T$_2$ CT absorption in triplet ESA (see Fig.~\ref{tripletPA}(b)), a result that is also true
for pentacene crystal \cite{Khan17a}. Finally, Fig.~\ref{TTPA}(b) shows triplet-triplet ESA in the mid-infrared region where there is no absorption from the free triplet. The final state of absorption here is the one-photon allowed S$_2$ state of
Fig.~\ref{grstabs}(a). Similar absorption cannot occur in the triplet subspace (see Fig.~\ref{tripletPA}(c) where we have shown the absence of triplet ESA in the mid-infrared).

{\it \underbar{Comparison to experiments and implication}.} We now point out the excellent semiquantitative 
agreements between the calculated versus published \cite{Sanders15a} and unpublished \cite{Sfeir} ground state and transient 
absorptions, 
and also make theoretical predictions. We then present the implications of the theoretical results for iSF.

The calculated ground state absorption spectra of Fig.~\ref{grstabs}(a) agree very well with the experimental spectra in Fig.~2 of Reference 
\onlinecite{Sanders15a}, except for the blue shift (instead of red shift) of the BP0 and BP1 absorptions to S$_1$, relative to the monomer absorption, and the relatively weak
intensity of the short wavelength absorption to S$_2$. We ascribe this to the limited active space in our MRSDCI calculations. 
It is well known that the one-photon optical states are predominantly 1e-1h with relatively little contribution from 2e-2h configurations 
\cite{Tavan87a,Chandross99a}. We speculate that the exclusion of the higher energy 1e-1h configurations in our choice of the active space is the reason behind 
these quantitative inaccuracies. This is confirmed from our singles-CI calculation reported in section III of the Supporting Information, where we
have retained all 1e-1h excitations but excluded ne-nh excitations with n$>$1.
As shown in Fig.~S3 in the 
Supporting Information, red shift of the dimer absorption as well as larger
intensity of the absorption to S$_2$ are both obtained now.
Our singles-CI calculation also uses the exciton basis, and the characterization of S$_2$ as a CT excitation remains valid.

The published experimental spin-singlet transient absorption spectrum of BP0 has shown PA in 450-550 nm region only (Fig.~3 in reference \onlinecite{Sanders15a}), corresponding to
our calculated singlet ESA in the 500-650 nm region in Fig.~\ref{singletPA}(a). Subsequent transient absorption measurements have found the PA corresponding to the
calculated ESA at $\sim 1200$ nm in Fig.~\ref{singletPA}(a) \cite{Sfeir}. {\it We predict additional PA in the mid infrared region beyond 1500 nm.} (see inset Fig.~5(a)). 
PA in this region has been found previously in $\pi$-conjugated polymers and single-walled carbon nanotubes \cite{Zhao06a}.

More interesting in the present context are the triplet and triplet-triplet ESAs. The calculated ESA spectrum of the triplet-triplet in BP0 is
shown in Figs.~8(a) and (b). Whether or not there is a difference between transient absorptions from the triplet and triplet-triplet
in materials exhibiting SF
has been a longstanding question. In pentacene crystal, the difference is subtle \cite{Herz15a,Pensack16a,Khan17a} because of the weak coupling between the monomers. There,
the ``intramonomer''  absorptions from T$_1$ and $^1$(TT)$_1$ occur at nearly the same wavelength, a feature that remains true in our calculations
for BP0 (see absorptions near $\sim$ 550 nm in Figs.~\ref{tripletPA}(b) and ~8(a)). Thus, based on the strong PAs in the visible alone, it is difficult to conclude whether the end product of the
photoexcitation is the triplet-triplet or free triplet. Figs.~3 and 5 of reference \onlinecite{Sanders15a}, however, find a second moderately strong
transient absorption in BP0 in the $\sim 675-775$ nm region in ``triplets obtained from singlet fission'', {\it but not in photosensitized triplets.} 
There occurs
a much weaker transient absorption at slightly {\it longer} wavelength in the photosensitized triplet \cite{Sanders15a}.
Based on this experimental information and
our computed ESA spectra in Figs.~6(b) and 8(a), we are led to believe that
primary photoproduct in the dimer is the triplet-triplet and not the free triplet. This feature, occurrence of a CT absorption in the triplet that is
weaker than the CT absorption in the triplet-triplet, at slightly
longer wavelength, is also true in the crystal \cite{Khan17a}. 

In addition to the absorption at the edge of the visible and the near IR, we have calculated an additional
ESA from $^1$(TT)$_1$ to S$_2$ in BP0, in the mid infrared region (see
Fig.~\ref{TTPA}(b)).{\it The counterpart of this ESA is absent in the free triplet T$_1$.} 
This PA has been experimentally observed in BP0 recently \cite{Sfeir}, albeit at slightly
shorter wavelength ($\sim 1200$ nm). The implication of this observation is clear: {\it direct photoexcitation of BPn is generating the $^1$(TT)$_1$ two-photon state and not free
triplets}. This does not preclude SF since triplets may be generated at longer times driven by intermolecular couplings. Our conclusion is in agreement with that reached 
by Sanders {\it et al.} in their more recent investigation of pentacene-tetracene heterodimers \cite{Sanders16a}. The most likely reason behind the occurrence of the 
calculated $^1$(TT)$_1 \to$ S$_2$ infrared absorption  at longer wavelength than in the experiment is that the true $^1$(TT)$_1$ occurs at an energy lower than calculated.
Recall that in the polyenes, the energy of the triplet-triplet 2$^1$A$_g^-$ decreases with length faster than the optical 1$^1$B$_u^+$ exciton \cite{Kohler88a}. 
Recall also that in pentacene monomer, the calculated $^1$(TT)$_1$ occurs slightly below S$_1$ \cite{Khan17a}. 
It is then likely that the true $^1$(TT)$_1$ occurs at an even lower energy than S1 in the dimer, even as our calculations find them to be quasidegenerate.
The quasidegeneracy of S$_1$ and $^1$(TT)$_1$ found in our calculation may be an artifact of the limited CI and the incomplete active space. As with the 2$^1$A$_g^-$, inclusion
of even higher order CI should lower the energy of the $^1$(TT)$_1$ further. 

{\it \underbar{Conclusion and Outlook}.} In summary, we have performed correlated-electron calculations that include CI with up to dominant 4e-4h excitations for the optical singlet,
lowest triplet and the triplet-triplet states in BPn, n = 0$-$2, using an exciton basis, over an active space of 24 MOs. 
We have also performed similar high order CI calculations of
ESAs from S$_1$, T$_1$ and $^1$(TT)$_1$ for BP0 over extended wavelength region, for comparison to published \cite{Sanders15a} and unpublished \cite{Sfeir}
experimental results. We are able to 
give physical intepretations of all excitations within a pictorial exciton basis desciption of eigenstates.
We find that the benzene MOs contribute very weakly to the S$_1$, T$_1$ and $^1$(TT)$_1$ wavefunctions in BP1 and BP2. 
We find singlet ESAs at wavelengths longer than the observed PA in the 500-600 nm range
in the published work \cite{Sanders15a}, in the 1000-1200 nm and at even longer wavelength ($> 2000$ nm). PA at $\sim 1000$ nm has been
observed recently \cite{Sfeir}. Most importantly, we find significant difference in the calculated ESAs from the T$_1$ and the $^1$(TT)$_1$ in BP0,
in contrast to pentacene crystal, where the difference is subtle. Based on the calculated difference, the published PA spectra \cite{Sanders15a},
and very recent observation of long wavelength ($> 1000$) nm PA from a {\it non-singlet} state \cite{Sfeir} we conclude that the primary product of photoexcitation in BPn
is $^1$(TT)$_1$ and not free triplets. In principle, strong interdimer coupling may lead to further dissociation of the $^1$(TT)$_1$ leading to the
generation of free triplets.

Finally, much of the theoretical work on both xSF and iSF until now has been performed using a very small active space and limited CI, 
sometimes including only the configurations shown in Fig.~3(b)
and a few others. 
Our calculations clearly show the need to have both a large active space and to perform high order CI calculations. To begin with, unbiased determination 
of $^1$(TT)$_1$ wavefunction is not possible without performing such calculations. As seen in Fig.~\ref{triplettriplet}(a), only about 80$\%$ of the true
$^1$(TT)$_1$ is a simple product of two triplets in the two pentacenes. More importantly, calculations of ESAs that allow us to distinguish between
T$_1$ and $^1$(TT)$_1$ are not possible without retaining a large active space. Actually, even our relatively large active space that retains 24 MOs 
is insufficient for obtaining the correct intensities of the absorption to S$_2$ in Fig.~3(a), or for performing ESA calculations for BP1 and BP2. 
High order CI calculations while retaining an even larger active space will be necessary for  obtaining ESA spectra of the latter. 

\appendix
\newcommand{\beginappendix}{%
        \setcounter{table}{0}
        \renewcommand{\thetable}{A.\Roman{table}}%
        \setcounter{figure}{0}
        \renewcommand{\thefigure}{A.\arabic{figure}}%
     }
\beginappendix
\section{Ground state absorption spectrum of BP2}
\vskip 0.5pc
\begin{table}[H]
\centering
\setlength{\tabcolsep}{8pt}
\caption{Calculated energies (in eV) of the three lowest optical, triplet and triplet-triplet states of BP2 and $U$ = 6.7 eV, $\kappa$ = 1.0.}
\begin{tabular}{l c c c c c}
\hline\hline
  Compound    &  S$_1$   &   S$_1^*$   &   S$_2$     &     T$_1$   &    $^1$(TT)$_1$ \\[0.5ex]
\hline

BP2 & 2.1 & 3.11 & 3.26 & 1.26 & 2.35 \\ [0.5ex]

\hline
\end{tabular}
\label{BP2}
\end{table}
\begin{figure}[H]
\includegraphics[width=3.5in]{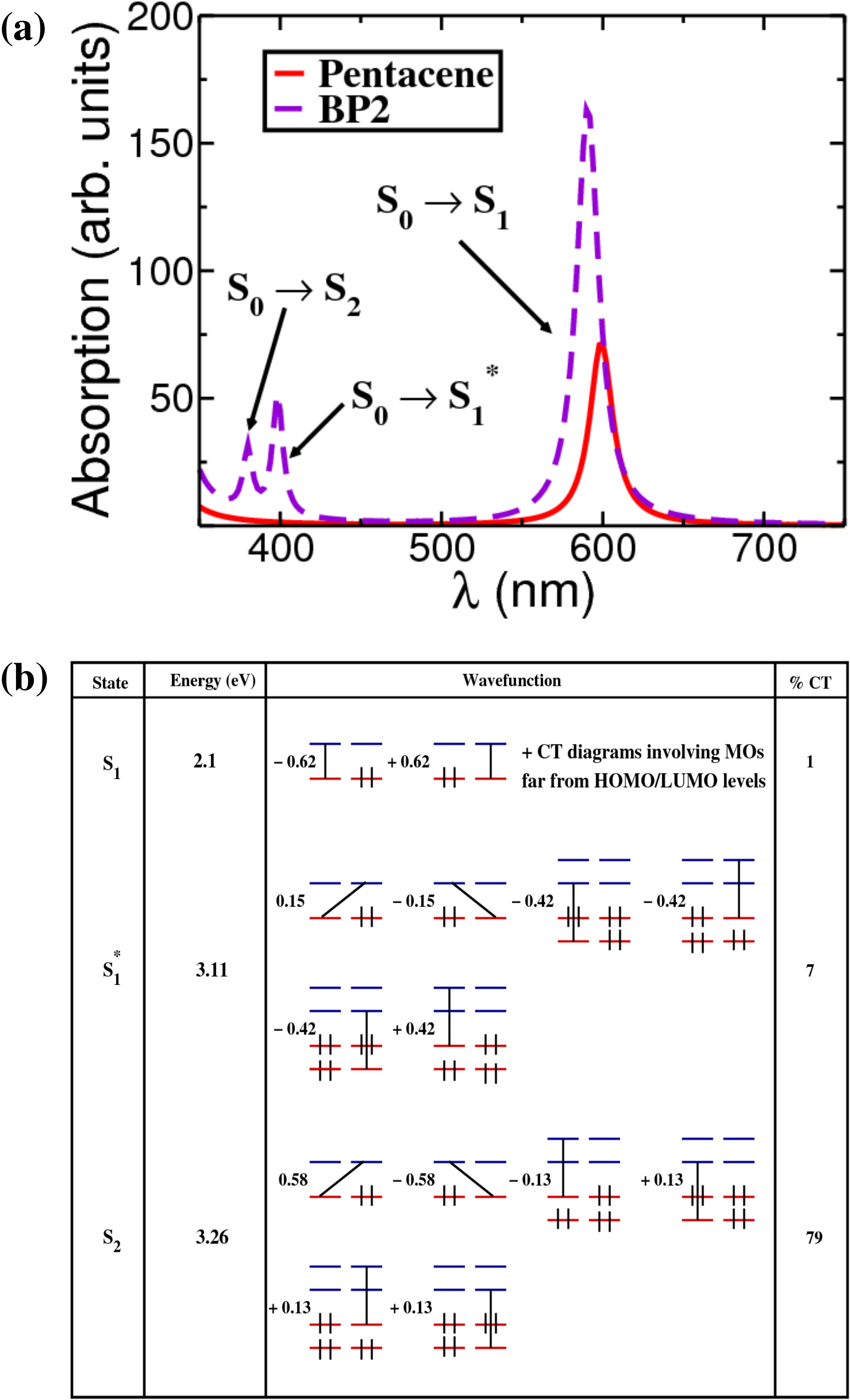}
\caption{(a) Ground state absorption spectrum of a pentacene monomer (solid red) and modified BP2 (dashed violet - Here, BP2 refers to two pentacene molecules covalently linked
with the help of two phenyl spacers) and (b) Dominant excitonic configurations to the final states in the absorption spectrum - S$_1$, S$_1^*$ and S$_2$.}
\label{BP2spectrum}
\end{figure}
Computational constraints prevent us from including the TIPS group in our calculations of BP2. Hence, we performed an MRSDCI calculation
of the ground state absorption spectrum of two pentacene molecules tethered to each other with the help of two phenyl linkers (modified BP2) as well as the monomer (pentacene) with the same set of parameters
as has been used for the other TIPS-dimers (BP0, BP1). On comparing the spectrum with that of BP0 and BP1, we notice that
the optical signals are not only blue-shifted
but the higher energy CT state (S$_2$) splits into two states (S$_1^*$ and S$_2$). Both these states have
 a weak dipole coupling with the ground state.
While the relative weights of the CT diagrams in S$_2$ are larger, S$_1$ with the same configurations as S$_2$ has strong contributions from intra-monomer transitions. 
It is therefore conceivable that with the increase in the separation of the pentacene units, the higher energy CT state would further split into a new Frenkel and CT state. 
It is apparent that the modified spectrum above
closely resembles the ground state absorption spectrum obtained for BP0 and BP1. As before, the benzene orbitals play an insignificant role in the description of the elctronic states in BP2.

\noindent {\bf Acknowledgments.} 
We acknowledge many helpful discussions with Professor 
Alok Shukla (IIT Bombay) and partial support from NSF-CHE-1151475 and Arizona TRIF photonics. We are thankful to Dr. Matthew Sfeir (Brookhaven 
National Laboratory) for sharing with us unpublished transient absorption data.

\end{document}